\documentclass[sn-mathphys]{sn-jnl}

\jyear{2022}%

\usepackage{cleveref}
\usepackage{caption} 
\theoremstyle{thmstyleone}%
\theoremstyle{thmstyletwo}%

\theoremstyle{thmstylethree}%

\begin{document}

\title[A Harmonized Requirements Quality Theory]{Requirements Quality Research: a harmonized Theory, Evaluation, and Roadmap\footnote{The version of record of this article, first published in \emph{Requirements Engineering}, is available online at Publisher’s website: \url{http://dx.doi.org/10.1007/s00766-023-00405-y}}}


\author*[1]{\fnm{Julian} \sur{Frattini}}\email{julian.frattini@bth.se}
\author[2]{\fnm{Lloyd} \sur{Montgomery}}\email{lloyd.montgomery@uni-hamburg.de}
\author[3,5]{\fnm{Jannik} \sur{Fischbach}}\email{jannik.fischbach@qualicen.de}

\author[1,5]{\fnm{Daniel} \sur{Mendez}}\email{daniel.mendez@bth.se}
\author[1]{\fnm{Davide} \sur{Fucci}}\email{davide.fucci@bth.se}
\author[1]{\fnm{Michael} \sur{Unterkalmsteiner}}\email{michael.unterkalmsteiner@bth.se}

\affil*[1]{
    \orgname{Blekinge Institute of Technology}, 
    \orgaddress{
        \street{Valhallvägen 1}, 
        \city{Karlskrona}, 
        \postcode{37140}, 
        \country{Sweden}
    }
}

\affil[2]{
    \orgname{University of Hamburg}, 
    \orgaddress{
        \city{Hamburg}, 
        \postcode{20146}, 
        \country{Germany}
    }
}

\affil[3]{
    \orgname{Netlight Consulting GmbH}, 
    \orgaddress{
        \street{Sternstr. 5}, 
        \city{Munich}, 
        \postcode{80538}, 
        \country{Germany}
    }}

\affil[5]{
    \orgname{fortiss GmbH}, 
    \orgaddress{
        \street{Guerickestraße 25}, 
        \city{Munich}, 
        \postcode{80805}, 
        \country{Germany}
    }
}


\abstract{
    High-quality requirements minimize the risk of propagating defects to later stages of the software development life cycle. Achieving a sufficient level of quality is a major goal of requirements engineering. This requires a clear definition and understanding of requirements quality. 
    Though recent publications make an effort at disentangling the complex concept of quality, the requirements quality research community lacks identity and clear structure which guides advances and puts new findings into an holistic perspective.
    In this research commentary we contribute (1) a harmonized requirements quality theory organizing its core concepts, (2) an evaluation of the current state of requirements quality research, and (3) a research roadmap to guide advancements in the field.
    We show that requirements quality research focuses on normative rules and mostly fails to connect requirements quality to its impact on subsequent software development activities, impeding the relevance of the research.
    Adherence to the proposed requirements quality theory and following the outlined roadmap will be a step towards amending this gap. 
}

\keywords{Requirements Quality, Theory, Survey}



\maketitle


\section{Introduction}
\label{sec:intro}

The empirical evidence of the impact of requirements engineering (RE) on the software development life cycle has shown that the quality of requirements artifacts and processes influences project success and budget adherence~\cite{fernandez2017naming,wagner2019status,damian2006empirical}. Moreover, the cost of defects introduced during the RE phase of a project is reported to scale exponentially the longer they remain undetected~\cite{boehm1988understanding}. This necessitates quality assurance techniques capable of detecting RE defects as soon and as reliably as possible.

Requirements quality research is dedicated to supporting the software engineering process with the means to evaluate and improve the quality of requirements, mainly focusing on requirements artifacts~\cite{mendez2019artefacts}. However, recent systematic investigations of requirements quality literature revealed a lack of rigor and relevance of these contributions~\cite{montgomery2021empirical,frattini2022live}. Moreover, the impact of the quality factors proposed in literature (i.e., requirements writing rules) remains largely unexplored in practice~\cite{frattini2022live}, hindering its adoption in industry~\cite{franch2020practitioners,berry2012case,femmer2018requirements,phalp2007assessing}.

Existing quality theories and frameworks are too abstract to guide requirements quality research at an operational level~\cite{lindland1994understanding,pohl1993three}. These theories often only divide quality into sub-categories without any means of applicability. In this paper, we argue for the need for a theoretical and operationalizable foundation of requirements quality research. We review the closely related software quality research and draw parallels to requirements quality research to consolidate a harmonized requirements quality theory. Additionally, we survey requirements quality literature with respect to the theory to reveal current shortcomings. Accordingly, we make the following contributions:
\begin{enumerate}
    \item A harmonized requirements quality theory serving as a theoretical foundation for requirements quality research.
    \item A survey of requirements quality research revealing if and how concepts of the theory are reported in the state of the art, but also emphasizing shortcomings.
    \item A consequent research roadmap aimed at mitigating these shortcomings.
\end{enumerate}

The rest of this manuscript is organized as follows: \Cref{sec:sqr} illustrates the evolution of software quality research and draws the parallel to requirements quality research. In \Cref{sec:model}, we derive a harmonized requirements quality theory from this comparison. This theory is used to evaluate the state of requirements quality research in \Cref{sec:state} and reveal current shortcomings. The consequent research roadmap to mitigate these shortcomings is presented in \Cref{sec:roadmap} before concluding in \Cref{sec:conclusion}.

\section{Software Quality Research}
\label{sec:sqr}

Software quality research follows a similar premise as requirements quality research. It is necessary to control the quality of software artifacts (e.g., source code) as it impacts the overall quality of the development life cycle and the final product. This premise aligns with the aim of requirements quality research. To show commonalities and differences between these two research fields, we review the evolution of software quality research in \Cref{sec:sqr:sq} and draw a parallel to requirements quality research in \Cref{sec:sqr:re}. We reach conclusions about the necessary direction the latter needs to take.

\subsection{Evolution of Software Quality Research}
\label{sec:sqr:sq}

Software quality research revolves around assessing the quality of software artifacts~\cite{broy2005holistic}. In the following, we describe the evolution of the field according to Broy et al.~\cite{broy2005holistic} and Deissenboeck et al.~\cite{deissenboeck2007activity}.

\paragraph{Guidelines and Metrics-based approaches}

Guidelines are the simplest approach for controlling the quality of software artifacts. For example, the Java coding conventions~\cite{king2021code} prescribe---among other suggestions---how to name and structure Java files. However, guidelines commonly fail to significantly impact software quality, likely because they lack the motivation for their relevance~\cite{broy2006demystifying}. For example, the aforementioned suggestions are justified because ``[c]ode conventions improve the readability of the software''~\cite{king2021code} without any empirical evidence of that claim. Furthermore, guideline conformance is difficult to assess and hence seldom done in practice~\cite{deissenboeck2007activity}. The latter shortcoming was addressed by introducing metrics-based approaches where metrics were devised to measure relevant attributes of software artifacts. Among others, \textit{lines of code}~\cite{albrecht1983software} and \textit{cyclomatic complexity}~\cite{mccabe1976complexity} were used to evaluate software quality automatically. Nevertheless, most metrics continue to lack justification of their relevance~\cite{broy2005holistic,rosenberg1997some,khoshgoftaar1990lines,shepperd1988critique}.

\paragraph{Quality Models}
To overcome the relevance shortcoming, quality models aggregated metrics into hierarchical trees of criteria~\cite{mccall1977factors,boehm1976quantitative}. The leaf nodes are specific enough to be operationalized as an evaluation metric, while the aggregation into higher-level quality characteristics provided the justification for their relevance. For example, low-level concepts such as \textit{structuredness} and \textit{conciseness} of code were justified by their aggregation to \textit{understandability} and \textit{maintainability}, which were widely accepted as relevant software quality characteristics~\cite{boehm1976quantitative}. However, hierarchical models suffered from unclear decomposition rules and constrained levels of granularity, which were either too abstract to be operationalized or too detailed, disconnecting the applicable metrics from their rationale~\cite{broy2005holistic,deissenboeck2007activity}. 

\paragraph{Quality Meta-Models} 
The popularity of quality models necessitated a structure for the proposed models~\cite{kitchenham1997squid}. Meta-models like the Goal Question Metric approach by Basili et al.~\cite{basili1994goal} and the factor-strategy quality meta model by Marinescu and Ratiu~\cite{marinescu2004quantifying} provide this overarching structure. Deissenboeck et al.~\cite{deissenboeck2009software} contribute the DAP classification for quality models, which categorizes the aim of a quality model to be to \textit{define} (D), \textit{assess} (A), or \textit{predict} (P). The publication further relates quality meta-models to quality models as the ``model of the constructs and rules needed to build specific quality models.''~\cite{deissenboeck2009software}.

\paragraph{Activity-based Quality Models}
In addition to the shortcomings that existing quality models continued to suffer, the elements populating these models were found to be heterogeneous~\cite{deissenboeck2007activity}---i.e., properties of a \textit{system} were mixed with properties of \textit{activities in which the system is used}. For example, the maintainability branch in the software quality characteristics tree by Boehm et al.~\cite{boehm1978merritt} contains both system properties like the \textit{structuredness} of a software artifact, but also attributes of activities in which these artifacts are used, like \textit{modifiability}. The latter describes the \textit{activity} of \textit{modifying} an artifact rather than a system property, despite the adjective's nominalization suggesting otherwise.

So far, no clear rule for distinguishing a system from an activity property has been proposed. We derived two heuristics from the implicit argumentation of previous publications~\cite{deissenboeck2007activity}. First, if a property involves an additional agent (e.g., \textit{testability} involves a \textit{test engineer}, \textit{modifiability} involves a \textit{modifier}, although not necessarily human), then it represents how the system is used---i.e., an activity property. The second heuristic comes in the form of a syntactical criterion:
\begin{itemize}
    \item Nominalized adjectives (e.g., structured-ness, concise-ness) tend to be \textbf{system properties}
    \item Nominalized verbs (e.g., modify-ability, access-ability, augment-ability) tend to be \textbf{activity properties}
\end{itemize}
Interpreting activity properties as system properties ignores an underlying impact relationship. For example, interpreting \textit{modifiability} as the \textit{system} property of how receptive it is to change omits that actual system properties (e.g., whether the system is digital or analog or who has writing access rights) \textit{impact} the ability of a stakeholder to modify the system, which is an activity property.

\begin{figure}[h]
    \centering
    \includegraphics[width=\textwidth]{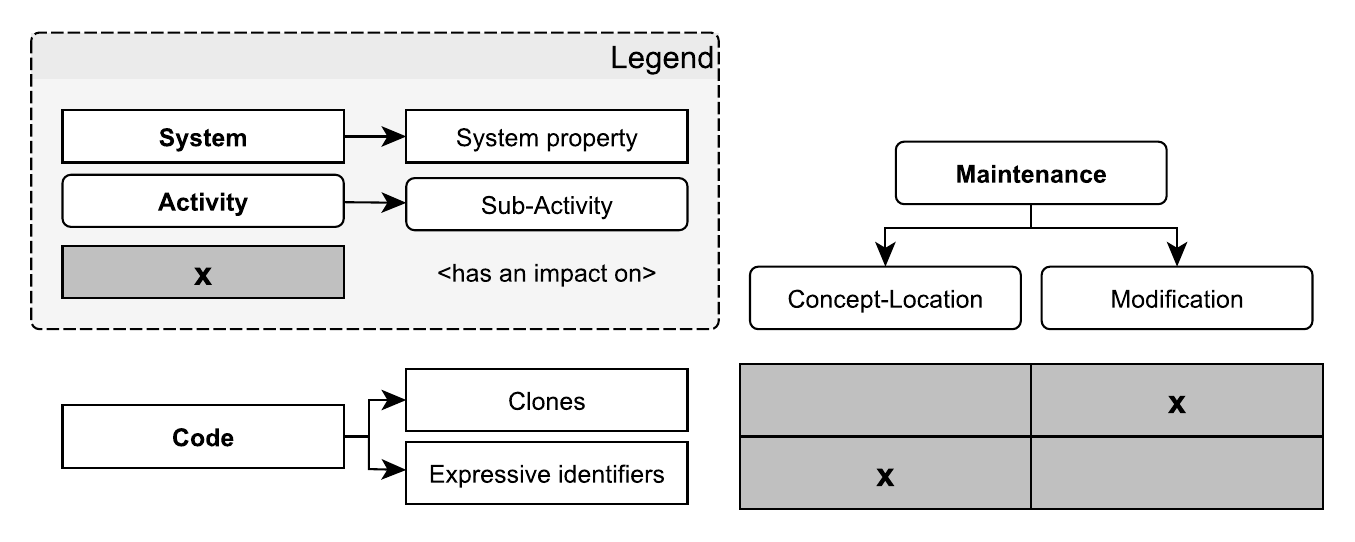}
    \caption{Excerpt from the activity-based quality model for maintainability}
    \label{fig:maintainability}
\end{figure}

To address the issue of heterogeneous properties, Deissenboeck et al. introduced \textit{activity-based quality models}~\cite{deissenboeck2007activity,broy2005holistic}, which separate system properties from activity properties and form two distinct, orthogonal dimensions. The model expresses quality as the impact of system properties on activity properties. \Cref{fig:maintainability} visualizes a simplified version of the quality model~\cite{deissenboeck2007activity}, showing how code clones impact the modification sub-activity and expressive identifiers impact the concept-location sub-activity.

The activity-based quality model was successfully applied to usability~\cite{winter2007comprehensive}, security~\cite{wagner2009security}, and service-oriented architecture~\cite{goeb2011software} before Wagner et al. distilled a comprehensive activity-based meta-model in the scope of the Quamoco project~\cite{wagner2012quamoco,wagner2012quamocotr}. In parallel, the original use case of the activity-based quality model, which focused on maintainability, received extensive tool support~\cite{deissenboeck2008tool,deissenboeck2011quamoco} contributing evidence to the operationalization of quality models in practice~\cite{steidl2014continuous}.

Activity-based quality models solve limitations of previous quality models at the cost of increased complexity, which manifests in additional challenges to operationalize and communicate the notion of quality~\cite{lochmann2013comprehensive}. However, the complexity of these models is necessary to tackle the faceted concept of quality~\cite{lochmann2013comprehensive,garvin1984really}. Research continuously tackles the inability of activity-based quality models to assess artifact quality and distinguish quality levels~\cite{klas2011evaluating}. For example, weights empirically derived from historical data replaced expert-based propositions~\cite{wagner2015operationalised}, and Bayesian networks were utilized to model the impact relationships~\cite{wagner2010bayesian}.

\subsection{Mapping to Requirements Quality Research}
\label{sec:sqr:re}

In the following, we draw a parallel of the evolution of quality research between the areas of software engineering and requirements engineering. 

\paragraph{Metrics and Quality Models}
Similar to software quality, requirements quality research historically originated from proposing metrics like \textit{passive voice} of requirements sentences~\cite{femmer2014impact} or \textit{sentence length}~\cite{ferrari2018detecting}, which are associated with bad quality of requirements specifications. Frattini et al.~\cite{frattini2022live} collected these quality factors and indicated their limitations. Most existing publications either fail to gauge the impact of these metrics~\cite{habib2021detecting} or explicitly disregard their relationship~\cite{femmer2017rapid}. Requirements quality models~\cite{berry2006new, lucassen2017improving} integrate these factors into larger frameworks but often remain vague on their notion of impact.

The investigation of impact is often limited to a comparison between the quality factor and practitioners' subjective, general perception of the quality of the requirements entities~\cite{parra2015methodology}. Wilson et al. contribute a first impact matrix between quality indicators and quality attributes~\cite{wilson1997automated}, but the latter suffers from the same system and activity properties heterogeneity. Similarly, Yang et al. state that ``[a]mbiguity is therefore not a property just of a text, but a conjoint property of the text and of the interpretations held by a group of readers of that text''~\cite{yang2011analysing}, exposing the necessary distinction between system and activity properties.

\paragraph{Activity-based Requirements Quality}
A large portion of requirements quality research exhibits the same shortcomings identified and overcome by software quality research, namely that (1) requirements quality factors lack relevance due to their unknown impact, which in turn inhibits adoption in practice, and (2) the terminology of requirements quality aspects confuses system and activity properties.

Femmer et al. apply the activity-based quality perspective to requirements engineering by proposing the activity-based requirements engineering quality model (ABRE-QM)~\cite{femmer2015ABREQM}. This model leverages the insights from activity-based software quality models~\cite{deissenboeck2007activity,broy2006demystifying,wagner2012quamoco} and shows that the quality of requirements depends on the impact they have on the activities in which they are used. However, despite the authors' call for action~\cite{femmer2018quality}, ABRE-QM saw little adoption in research as demonstrated in recent systematic investigations of the requirements quality literature~\cite{montgomery2021empirical,frattini2022live}. 

The ABRE-QM example above raises the concern that requirements quality researchers do not properly utilize the activity-based approach successfully employed in software quality research. In this manuscript, we want to encourage further research on this approach by presenting a revised requirements quality theory, a thorough investigation of the requirements quality literature verifying the hypotheses from previous studies~\cite{montgomery2021empirical,frattini2022live}, and a consequent research roadmap.

\section{Requirements Quality Theory}
\label{sec:model}

We generated a harmonized requirements quality theory (RQT) by consolidating the evolution of software quality models described in \Cref{sec:sqr:sq}, their application in requirements engineering as described in \Cref{sec:sqr:re}, and alignment to the established Quamoco quality model~\cite{wagner2012quamocotr}. In terms of theory types~\cite{gregor2006nature}, the RQT is both \textit{explanatory}, as it explains the notion of requirements quality, and \textit{prescriptive}, as it prescribes how to report contributions to requirements quality. The building blocks of the theory are described in \Cref{sec:model:meta} and illustrated with an example in \Cref{sec:model:example}.

\subsection{Theory}
\label{sec:model:meta}

\begin{figure}[ht]
    \centering
    \includegraphics[width=\textwidth]{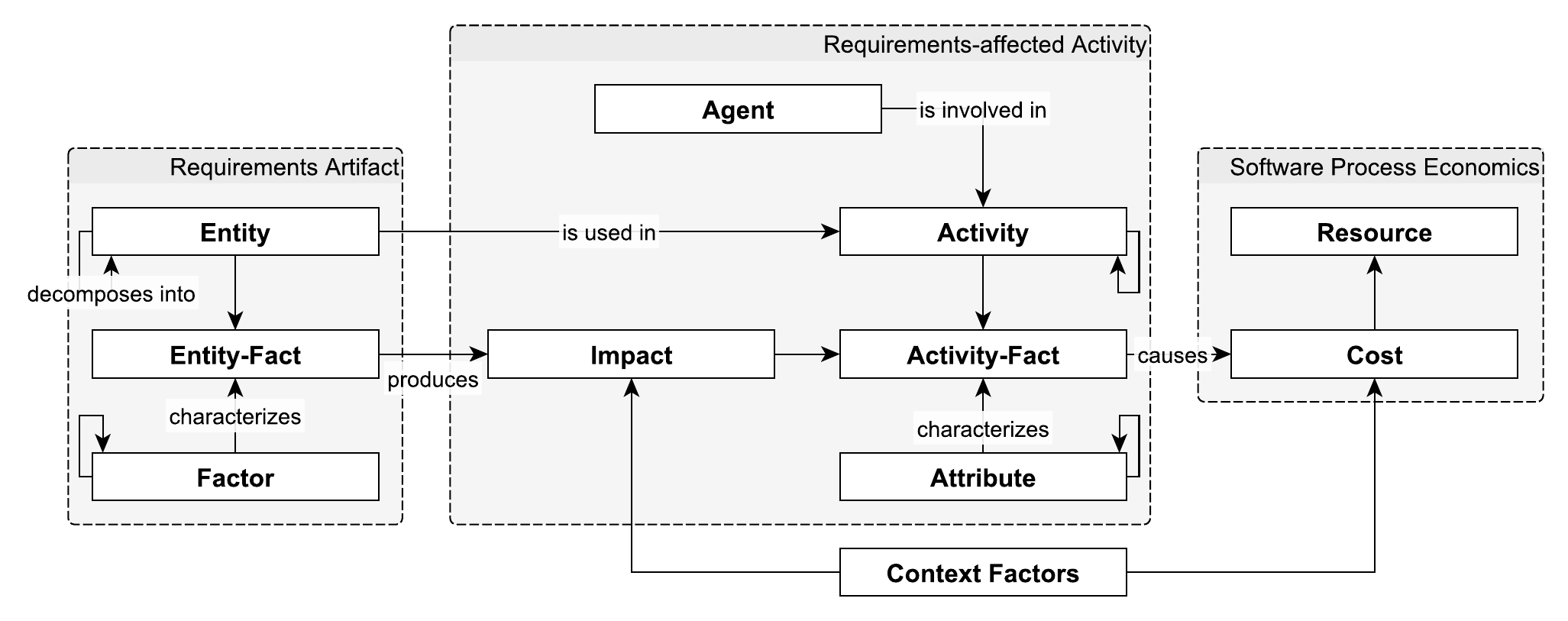}
    \caption{Concepts of the Requirements Quality Theory.}
    \label{fig:rqt:model}
\end{figure}

The concepts that constitute this theory are visualized in \Cref{fig:rqt:model}, and each concept is described in \Cref{tab:theory:concepts}. The model represents an evolution of the original activity-based requirements engineering quality model (ABRE-QM) proposed by Femmer et al.~\cite{femmer2015ABREQM}. Here, we present changes to the original model.

\begin{table}[ht]
    \centering
    \begin{tabular}{p{2cm}|p{7cm}|p{1cm}}
        \textbf{Concept} & \textbf{Explanation} & \textbf{Origin} \\ \toprule
        Entity & A requirements artifact or part thereof & \cite{femmer2015ABREQM} \\
        Factor & ``[A] normative metric which maps a textual requirement of a specific granularity''~\cite{frattini2022live} to a numerical output & \cite{deissenboeck2007activity,femmer2015ABREQM} \\
        Entity-Fact & A composition of one entity and one factor & \cite{deissenboeck2007activity} \\ \midrule
        Agent & Any person, group of people, or automatism involved in an activity & \cite{femmer2015ABREQM} \\
        Activity & An activity in which the entity is used & \cite{deissenboeck2007activity} \\
        Attribute & A measurable property of an activity & \cite{winter2007comprehensive} \\
        Activity-Fact & A composition of one activity and one attribute & \\
        Impact & The impact of a fact on an activity-fact & \cite{deissenboeck2007activity,femmer2015ABREQM} \\ \midrule
        Context Factor & A factor describing the context of the impact relationship & \cite{mund2015does,juergens2010much} \\ \midrule
        Cost & The magnitude of cost associated with an activity-fact & \cite{juergens2010much} \\
        Resource & The resource affected by the economical impact & \cite{deissenboeck2007economic,juergens2010much} \\ \bottomrule
    \end{tabular}
    \caption{Explanation and origin of theory concepts.}
    \label{tab:theory:concepts}
\end{table}

The artifact-related section of the model (left part of \Cref{fig:rqt:model}) is largely equivalent to the original publications~\cite{deissenboeck2007activity,femmer2015ABREQM}. Entities represent requirements artifacts of different granularity~\cite{mendez2019artefacts}, which can be decomposed into further entities. For example, a requirements specification can be decomposed into sections, which in turn consist of paragraphs and sentences or requirements. We consider an artifact to be a high-level requirements entity and hence do not explicitly add the \textit{artifact} to the model, deviating from the original~\cite{femmer2015ABREQM}. Similarly, factors can be decomposed into sub-factors to accommodate composite factors. For example, Antinyan et al.~\cite{antinyan2016complexity} position their proposed quality factor of \textit{conjunctive complexity} as a sub-factor of \textit{syntactical complexity}.

The activity-related section of the model (middle part of \Cref{fig:rqt:model}) again adapts the original models~\cite{deissenboeck2007activity,femmer2015ABREQM}. The concept \textit{activity} does not represent common requirements activities, like elicitation, analysis, and validation~\cite{sommerville2005integrated}, but rather every process that takes a requirements entity as input and produces an output. This includes some requirements activities (like analysis and validation, which use requirements as input) but not others (like elicitation, which often does not presuppose existing requirements). Hence, we rather refer to them as \textit{requirements-affected activities}. These further include implicit sub-activities (e.g., \textit{understanding} and \textit{interpreting} an entity), which can be aggregated with other, more explicit sub-activities (e.g., \textit{test case design}) to form high-level activities (e.g., \textit{validation}). The decomposition relationship of the activity concept accommodates this aggregation. To accommodate not only human actors involved in activities but also any automatism like requirements processing tools~\cite{fischbach2023automatic} we abstract the concept of \textit{stakeholder} to \textit{agent}.

We generalized the impact concept in this theory. While previous models assumed that impact is categorical (i.e., the occurrence of a fact has either a positive, negative, or no impact at all, like in \Cref{fig:maintainability}~\cite{deissenboeck2007activity} or linear (i.e., the larger the evaluation of a quality factor, the better/worse is its quality), we consider the impact to model any kind of relationship between Entity-facts and Activity-facts. This opens up the theory to more complex relationships, which can model the actual impact more accurately and allows to compare the impact of quality factors with each other.

Two concepts were added to the model. First, the impact was related to an \textit{Activity-fact} composed of an activity and an attribute as proposed by Winter et al.~\cite{winter2007comprehensive}. This way, the structure of the variables on the two sides of the impact relationship is mirrored. Furthermore, the necessity to associate an impact with a measurable property of an activity is emphasized. Second, context factors also influence the impact of an Entity-fact on an Activity-fact. As recognized by previous publications~\cite{mund2015does,juergens2010much}, the impact differs depending on external factors related to, among others, the organization and the people involved~\cite{petersen2009context}.

The economic section of the model (right part of \Cref{fig:rqt:model}) is a novel addition to previous iterations of the activity-based models~\cite{deissenboeck2007activity,femmer2015ABREQM,wagner2012quamocotr}. As long as the subsequent \textit{economic} impact of an Activity-fact is unknown, the Entity-fact that produces the Impact on this Activity-fact will remain neglected~\cite{deissenboeck2007economic,juergens2010much}. Hence, the software process economics perspective introduces a \textit{Cost} for a specific \textit{Resource} such as time or money.

\subsection{Example}
\label{sec:model:example}

\begin{figure}[h]
    \centering
    \includegraphics[width=\textwidth]{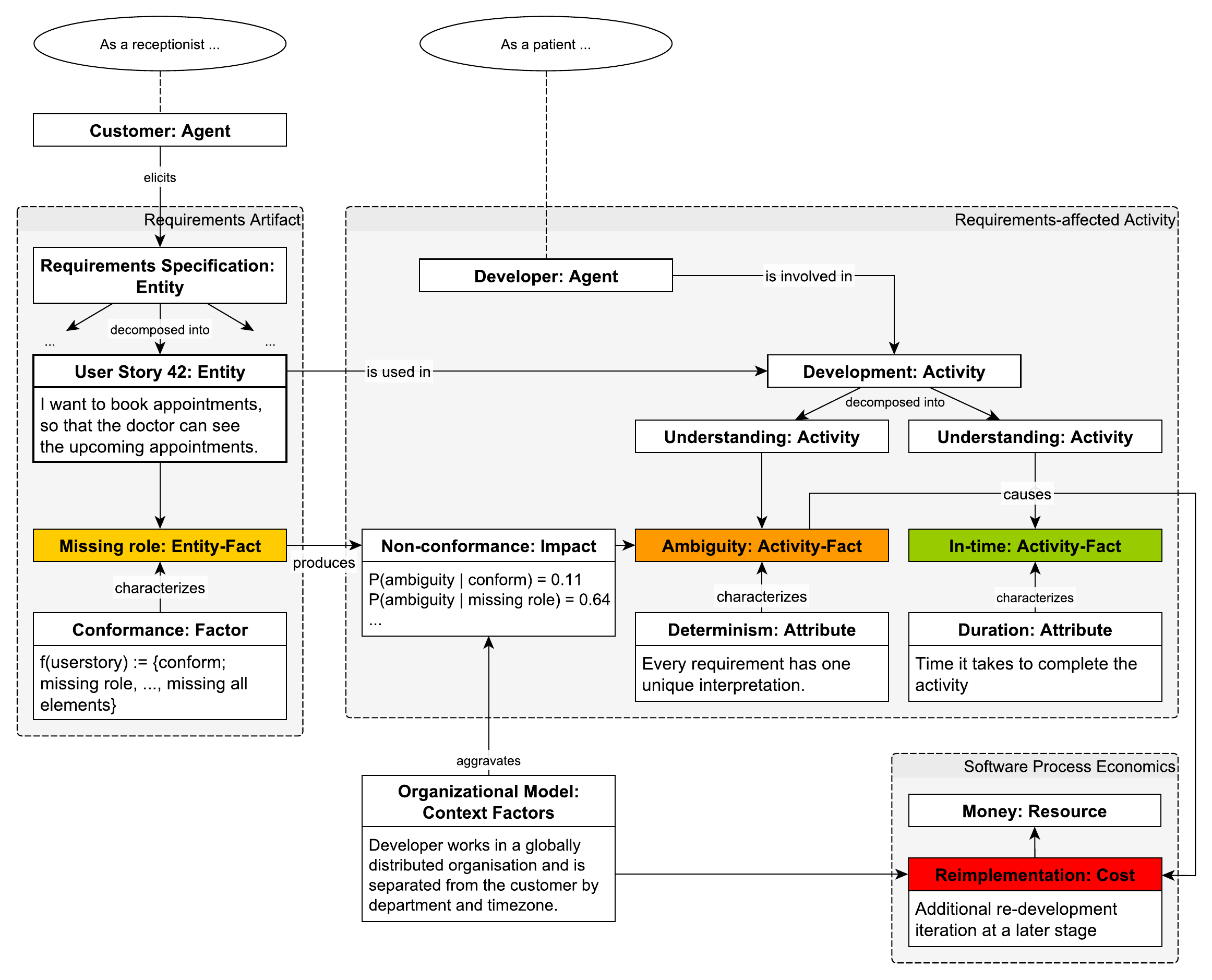}
    \caption{Exemplary instantiation of the theory}
    \label{fig:rqt:example}
\end{figure}

In this section, we illustrate the RQT with a fictitious example to demonstrate its application. The example is additionally visualized in \Cref{fig:rqt:example}.

In this example, a customer's requirements were elicited and documented in a requirements specification containing the entity \textit{user story 42}. One relevant quality factor used by the organization responsible for implementing the requirements is template \textit{conformance}, which prescribes that all user stories must follow the Connextra template~\cite{cohn2004user} ``As a $<$role$>$ I want to $<$goal$>$ so that $<$benefit$>$.'' This quality factor maps the entity to a categorical value, containing---among others---the values \textit{conform}, \textit{missing role}, and \textit{missing all elements}. In this example, the role is omitted from the user story. Hence, the quality factor template conformance is evaluated to \textit{missing role}, which constitutes the entity-fact (yellow box in \Cref{fig:rqt:example}).

The organization uses this user story in a subsequent, requirements-affected \textit{development} activity, where a different stakeholder---the developer---is responsible for translating the entity into code. This activity can be decomposed into two distinct sub-activities: \textit{understanding} the entity and \textit{programming} the respective implementation.

One desired attribute of the activity understanding is \textit{determinism}---i.e., a requirements entity should have only one unique interpretation. Possible variations of the interpretation and, therefore, the subsequent translation of a requirement must be avoided. Because the \textit{conformance} quality factor is evaluated to \textit{missing role} on the \textit{user story} entity, the \textit{understanding} activity is less \textit{deterministic}, as the developer can make a different assumption about the role implied by the requirement. The understanding activity has become ambiguous, which constitutes the \textit{activity-fact} (orange box in \Cref{fig:rqt:example}). 

The relationship between the entity-fact and the activity-fact is the \textit{impact} of the quality factor. Instead of limiting the impact concept to categorical values (e.g., either \textit{has an impact} or \textit{has no impact}), the RQT enables more complex impact relationships. In this fictitious example, the quality factor value \textit{missing role} is associated with a 64\% chance of making the understanding sub-activity ambiguous. This relationship can be determined empirically via experimental research investigating the likelihood of the different values of the conformance quality factor reducing the determinism of the understanding sub-activity.

The programming sub-activity may go unaffected by the entity-fact that the conformance has a value of \textit{missing role} (green box in \Cref{fig:rqt:example}): regardless of the agent's interpretation of the requirements entity, the programming sub-activity will remain unaffected in respect to the relevant attribute \textit{duration} under the assumption of a similar user interface for both roles. Whether the feature is coded for the role receptionist (as the customer intended) or patient (as the developer assumed) does not significantly change the duration of the sub-activity if the user interfaces only barely differ.

The significant impact on understanding is influenced by the organizational model, which is one relevant \textit{context factor}. Since the organization is globally distributed and the two involved agents are unlikely to have informal interactions, the impact is amplified. In contrast, in a small organization where all involved agents share an office, the impact can be alleviated as missing information is recovered through informal communication. Similarly, the software development process model may significantly influence the impact of the quality factor, and the use of an agile approach may reduce the impact by encouraging communication between the customer and developer. The context factors significantly influence the impact and, therefore, have to be included in the relationship between entity-facts and activity-facts.

The reduced determinism of the understanding activity has an economic effect---i.e., the less deterministic the activity is, the more the implementation needs to be revised, which costs money and time (red box in \Cref{fig:rqt:example}). Context factors influence the extent of this effect as, for example, a re-implementation can be more costly in larger organizations due to organizational overhead.

For the sake of brevity, the example omits the following aspects: (1) the example limits the number of elements populating the relationship. More quality factors of the entity, activities, attributes of activities, and context factors are possibly involved in the relationship. (2) Interaction effects between quality factors and context factors are plausible but not reported here.

However, the example demonstrates how adherence to this activity-based RQT elevates requirements quality factors from normative rules (i.e., user stories must conform the template for the sake of it) to empirically-backed impact predictions (i.e., user stories must conform the template to mitigate ambiguous interpretations and avoid implementation cost).

\section{State of research}
\label{sec:state}

Despite the publication of the ABRE-QM~\cite{femmer2015ABREQM} and its authors' proposition to adapt the quality meta-model for future requirements quality research~\cite{femmer2018quality}, recent systematic reviews raised concerns regarding a perspective on requirements quality limited to the artifact-related section of the model (left part of \Cref{fig:rqt:model})~\cite{montgomery2021empirical,frattini2022live}.

To validate these concerns, we formulate the following research question. \textbf{How are the concepts of the requirements quality theory reported in requirements quality literature?} Answering this research question requires extracting information from a population of publications; accordingly, we employ survey research as our approach to gain insight into the current state of research. We follow the survey guidelines by Moll{\'e}ri et al.~\cite{molleri2020empirically} and report our survey in the following subsections. All supplementary material for replicating this study is available in our replication package\footnote{Available at \url{https://doi.org/10.5281/zenodo.8167598}.}.

\subsection{Survey Objects}
\label{sec:state:objects}

The target population of our survey is the requirements quality literature dealing with quality factors in requirements artifacts. Frattini et al.~\cite{frattini2022live} conducted a systematic study on requirements quality factors, including a sample of 57 primary studies. To our knowledge, this is the only sample that fulfills our aforementioned requirements. This classifies the sampling as non-probabilistic, more specifically convenience sampling~\cite{molleri2020empirically}.

\subsection{Study Design}
\label{sec:state:design}

We follow the recommended practices for the survey research process and report our steps accordingly~\cite{molleri2020empirically}. However, we disregarded steps that only apply to surveys with human subjects, such as \textit{participant recruitment} and \textit{response management}.

We derived the \textit{definition of the research objectives} in the form of the research question directly from previous research~\cite{femmer2018quality,montgomery2021empirical,frattini2022live}. We established a \textit{study plan}, rigorously documenting all research progress and justifications for any deviations during the process. We \textit{identified and characterized the population} of our survey and executed our \textit{sampling plan} as described in~\Cref{sec:state:objects}. 

For our \textit{instrument design}, we maintained two artifacts. We created an extraction guideline based on the RQT concepts. Each concept of the RQT was associated with one or more categorical variables, each containing a set of codes that represented \textit{if} and \textit{how} the concept was reported. The codes were created ad hoc in the first iteration of extraction and refined based on discussions and theoretical background in the second iteration.

The extent of the codes varied. The codes that represent how the concept \textit{entity} is reported are, for example, \textit{explicit} and \textit{implicit}. An entity is either reported explicitly if its scope and form are clear. It is reported implicitly if the authors just report that the factor applies to a ``requirement'' without defining whether this is a single or multiple natural language sentence, whether the language is constrained or not, or whether it assumes a full sentence at all. 

The codes of other concepts were more complex and grouped into distinct categories. For example, the codes of the concept \textit{Factor} were split into two groups, representing both the \textit{explicitness} when reporting a factor (i.e., whether the factor is explicitly \textit{reported} or \textit{referenced} from another publication) and the \textit{form} in which the factor is reported (i.e. if the factor is represented with a \textit{textual description} or defined using a logical or mathematical \textit{formula}). The extraction guideline containing all codes, explanations, and examples can be found in the replication package.

The first author extracted the appropriate code for each concept in the requirements quality theory from each publication. The extractions for each publication in the sample were recorded in a spreadsheet. For \textit{instrument validation}, the second author of this manuscript independently performed the extraction task using the guideline on six ($\approx10\%$) publications randomly sampled from the survey objects. The second author performed the extraction on two of these six publications as training, and the remaining four were used to calculate the inter-rater reliability between the first and second author.

The task overlap achieved an percentage agreement~\cite{holsti1969content} of $83.3\%$, whereas Cohen's Kappa yields a \textit{moderate} agreement of $54.2\%$. As Cohen's Kappa is unreliable for uneven marginal distributions~\cite{feng2015mistakes}, we calculated the more robust S-Score~\cite{bennett1954communications}---yielding a \textit{good} agreement of $76.8\%$---which we deem sufficient for assessing the inter-rater reliability.

We used the codes in the \textit{data analysis} phase to generate descriptive statistics on which we based our interpretation of the state of requirements quality. These form a quantified foundation for interpreting the state of requirements quality literature with respect to the research question. For final \textit{reporting}, we adapted established reporting guidelines~\cite{molleri2020empirically} and disclosed all material in a reusable replication package.

\subsection{Study Results}
\label{sec:state:results}

\begin{figure}
    \centering
    \includegraphics[width=\textwidth]{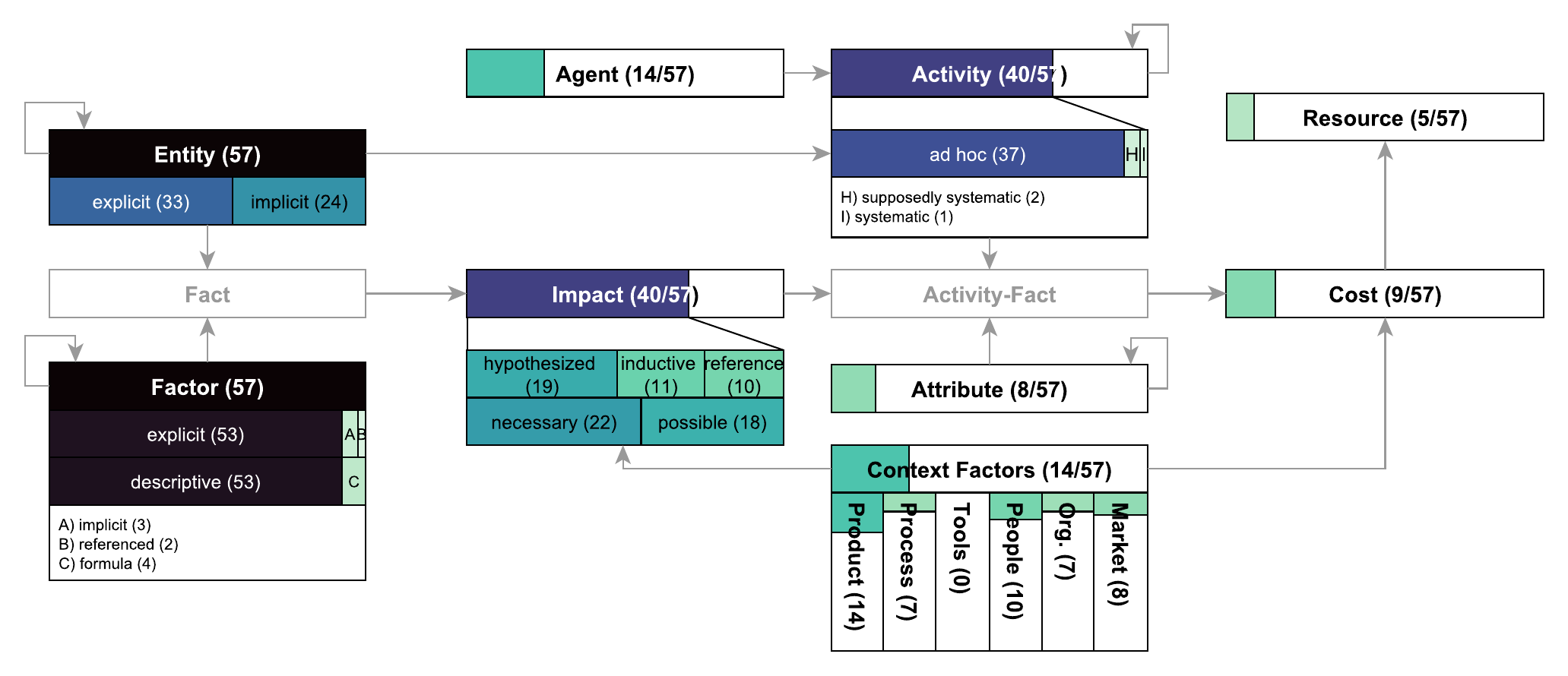}
    \includegraphics[width=0.8\textwidth]{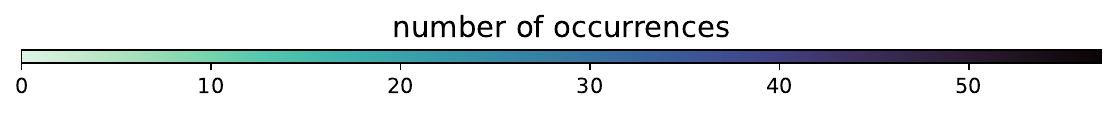}
    \caption{Survey results depicting the distribution of codes.}
    \label{fig:rqt:results}
\end{figure}

\Cref{fig:rqt:results} visualizes the distribution of the relevant codes among all concepts included in the requirements quality theory. Each concept is overlaid with a bar representing how many of the 57 publications contained the concept. The row below each concept represents its dimensions derived from the appropriate codes.

Though both entities and factors are explicitly reported in all 57 publications of the sample, a large portion ($24/57 = 42.1\%$) of entities is reported implicitly---i.e., the entity's scope is not clear. This occurs mostly because authors attach the reported quality factor to the entity \textit{requirement} without specifying the scope or form of the entity. Montgomery et al.~\cite{montgomery2021empirical} have already noted this shortcoming in the requirements quality literature and it represents a terminological ambiguity in the research domain.

Seventeen out of 57 publications ($29.8\%$) do not report any impact on activities (code \textit{N/A}) and hence neglect the practical relevance of the proposed quality factors. Agents are only reported in 14 ($24.6\%$) of all publications. Activities are---when reported---predominantly elicited \textit{ad hoc} ($37/40 = 92\%$) and rarely \textit{systematically}---i.e., when activities impacted by a quality factor are discussed, the identification of activities has no systematic approach. Attributes are also only rarely reported ($8/57 = 14\%$).

We grouped the codes classifying how \textit{impact} is reported into four distinct dimensions, two of which are reported here. The \textit{evidence} for the impact---when at all reported---is dominantly hypothesized ($19/40 = 47.5\%$) and rarely either inductive ($11/40 = 27.5\%$) or referenced ($10/40 = 25\%$), i.e., draws the evidence from another publication. Previous studies~\cite{montgomery2021empirical,frattini2022live} have also noted this dominance of anecdotal, non-empirical evidence. The \textit{modality} of impact relationships is balanced between \textit{necessary} and \textit{possible}---i.e., the impact of quality factors is reported almost equally often to be certain or potential. The remaining two dimensions of impact (\textit{generality} and \textit{frame of reference}) yielded no additional insight into the surveyed objects and are hence not reported here but contained in the replication package.


Context factors are almost completely neglected and only reported to a degree varying between zero (no publication reports the influence of any \textit{tools}) and $24.6\%$ (14 out of 57 publications reporting \textit{product}-related factors, e.g., the system's size or type).

Both \textit{cost} and \textit{resources} are reported only rarely ($9/57 = 15.8\%$ and $5/57 = 8.8\%$ respectively) and, if so, only hypothesized or referenced, never determined empirically. Money and time are mentioned as the resources affected by activity impact, and the cost is only estimated in terms of expected change (e.g., ``\textit{reduction} of the time spent''~\cite{femmer2017rapid}) or general magnitude (e.g., ``\textit{significant} amounts of money''~\cite{femmer2017requirements}).

\subsection{Interpretation}
\label{sec:state:interpretation}

In this section, we interpret the results presented in \Cref{sec:state:results} and answer the research question.

Publications in the requirements quality literature adhere to the RQT to a varying degree. While all publications in the sample mentioned both an entity and a quality factor, activity-related concepts, context factors, and the economic impact are often neglected. Failing to consider the context factors severely threatens the external validity of the proposed quality factors~\cite{mund2015does,juergens2010much} and neglecting the economic impact risks undermines their acceptance~\cite{deissenboeck2007economic,juergens2010much}.

Context factors and economic impact are arguably more challenging to investigate~\cite{kamata2007does}; however, we emphasize that the lack of activity perspective when proposing quality factors is critical for several reasons. The complete negligence of a quality factor's impact limits the factor to a normative, unmotivated prescription and challenges its practical relevance~\cite{femmer2015ABREQM}, which in turn promotes skepticism regarding requirements quality factors in industry~\cite{franch2020practitioners,berry2012case,femmer2018requirements,phalp2007assessing}. 

The survey emphasized two additional shortcomings in the field of requirements quality research. First, the tendency to elicit activities \textit{ad hoc} when discussing the impact of requirements quality factors bears the risk of missing other important impacts. Most publications discuss a hypothesized impact of a quality factor on a non-systematically selected activity or set of activities. This selection is usually justified by anecdotal or folkloric circumstances, like ``[a]mbiguous requirements may bring about misinterpretations among stakeholders, and prompt a few issues''~\cite{sinpang2017detecting}. 

While these impact relationships are neither empirically proven nor falsified, the non-systematic selection of activities can disregard other impact relationships. Femmer et al.~\cite{femmer2015ABREQM} demonstrated that a systematic elicitation of activities could reveal both positive and negative impacts by the same quality factor. For example, the factor \textit{free of UI design details}, which states that an ``artifact should describe the problem domain instead of the solution domain''~\cite{femmer2015ABREQM}, will positively affect maintainability, as UI details are volatile in the beginning and require a lot of change management if specified in a requirement. Conversely, the same factor negatively impacts understandability, as the presence of UI design makes requirements more comprehensible.

Second, while activities are not reported consistently, attributes of activities are reported even less. Attributes represent measurable characteristics of activities; for example, the activity \textit{understanding} can be quantified by its attribute \textit{level of agreement}~\cite{antinyan2016complexity,chantree2006identifying} or a \textit{readability index}~\cite{din2008requirements}. Neglecting the quantifiable attributes of activities impedes an empirical evaluation of a quality factor impact because it omits the measurement instrument for the dependent variable (i.e., the activity-fact) in the impact relationship~\cite{winter2007comprehensive}.

We conclude that the requirements quality theory is implicitly embedded in the requirements quality literature. However, insufficient adherence to it results in several limitations when reporting new requirements quality factors. While the artifact-centric theory concepts are commonly covered, activity-centric concepts, context factors, and economic concepts receive less attention, which decreases these publications' practical relevance. With this study, we empirically confirm the concerns voiced in previous investigations of the requirements quality literature~\cite{montgomery2021empirical,frattini2022live}.

\subsection{Threats to Validity of this Research}
\label{sec:state:validity}

We discuss the threats to validity proposed by Wohlin et al.~\cite{wohlin2012experimentation} and extended by Moll{\'e}ri et al.~\cite{molleri2020empirically}.

\paragraph{Internal Validity} We acknowledge a threat to internal validity due to sampling of publications. The method of object selection~\cite{montgomery2021empirical,frattini2022live} is deemed sufficiently rigorous to derive an initial theory.

\paragraph{Construct Validity} The constructs in this study---i.e., the elements of the theory---are established strictly following mature quality theories from the field of software quality. This ensures the alignment between the underlying theory and measurement constructs.

The lack of a theory to which the surveyed publications could have adhered when reporting quality factors resulted in the concepts of requirements quality often being embedded implicitly, complicating the extraction task. We minimized the resulting threat to internal validity through independent labeling and calculating appropriate inter-rater reliability metrics~\cite{feng2015mistakes}.

\paragraph{External Validity} The selected sample of publications~\cite{frattini2022live} is constrained to empirical contributions to requirements quality research~\cite{montgomery2021empirical}. This limits the conclusion validity of the type of evidence for the \textit{impact} concept, as non-empirical work could contribute \textit{theoretical} evidence for impact relationships. For example, the impact of quality factors like \textit{nominalization}~\cite{landhausser2015denom} can be derived deductively by referring to valency reduction caused by nominalization~\cite{mackenzie1985nominalization}. While publications utilizing linguistic theory are unknown to the authors, a valid conclusion regarding this type of evidence requires a more thorough extension of the sampling strategy.

\section{Research Roadmap}
\label{sec:roadmap}

Femmer et al. proposed an initial research roadmap detailing how to advance the field of requirements quality research~\cite{femmer2018quality}. Based on concerns of previous studies~\cite{montgomery2021empirical,frattini2022live} and the survey of the state of research reported in this study, we assess and update the three suggested steps by Femmer et al.~\cite{femmer2018quality}:
\begin{enumerate}
    \item Creation of ``a reference artifact and a usage model'' eliciting typical entities, activities, and agents.
    \item Creation of ``a taxonomy of quality factors'' as a central, accessible repository of quality factors.
    \item Creation of ``a taxonomy of impacts'' as a catalog of impacts from quality factors onto activities.
\end{enumerate}
We reflect on these proposed research streams in \Cref{sec:roadmap:artifacts,sec:roadmap:factors,sec:roadmap:impact} and add three further proposals in \Cref{sec:roadmap:context,sec:roadmap:economics,sec:roadmap:tool}. Because these research streams are grounded in the experiences from the software quality research, we expect contributions to them to promote requirements quality research that is relevant to practice.

\subsection{Artifact and Usage Model}
\label{sec:roadmap:artifacts}
Mendez et al. have contributed a reference artifact model for requirements engineering~\cite{mendez2019artefacts,mendez2010meta} based on their fundamental positioning on artifact orientation~\cite{mendez2015artefact,mendez2013improving}. The AMDiRE approach constitutes a domain-agnostic reference for artifact types and serves the purpose requested by Femmer et al.~\cite{femmer2018quality} in that it can be tailored towards any industry context to model an artifact structure. 

While the elicitation of human~\cite{sharp1999stakeholder} and non-human, automatic agents~\cite{zhao2021natural} has been addressed, a reference model for activities requires explicit attention in literature. More importantly, with the update of the requirements quality theory over the initial ABRE-QM~\cite{femmer2015ABREQM}, we argue that a reference model for requirements-affected activities needs to provide \textit{attributes} to quantify each activity. Such attributes enable an empirical assessment of the impact of quality factors.

Additionally, a majority of publications reporting an impacted activity mention some variation of \textit{understanding} or \textit{interpreting} ($32/40 = 80\%$). We assume that every requirements-affected activity comprises an initial \textit{interpretation} sub-activity. However, such composition is obscured by the lack of a proper reference model for requirements-affected activities accounting for their aggregated nature.

It is conceivable that the \textit{interpretation} sub-activity is most prone to defects, which explains the research community's focus on \textit{ambiguity}~\cite{montgomery2021empirical}, as ambiguity represents the non-determinism of an interpretation. We argue that a proper reference model for requirements-affected activities accounting for their aggregated nature can steer research towards identifying critical sub-activities---i.e., the ones most prone to impacting subsequent activities. 

\subsection{Taxonomy of Quality Factors}
\label{sec:roadmap:factors}
Requirements quality factors~\cite{femmer2018quality,frattini2022live} are the cornerstone of artifact-centric quality assurance. The requirements quality factor ontology proposed by Frattini et al.~\cite{frattini2022live} furthered this research stream. Although the ontology is in an early stage and requires additional iterations, quality factors and related objects---such as data sets and automation approaches---are now collected in a central repository.

\subsection{Taxonomy of Impacts}
\label{sec:roadmap:impact}
The taxonomy of impacts that Femmer et al.~\cite{femmer2018quality} deem the necessary final step of the roadmap has to be extended. Previous quality models---including the ABRE-QM~\cite{femmer2015ABREQM}---consider only categorical or, at most, linear impact relationships. Therefore, a taxonomy seemed sufficient to record ``a list of well-examined effects of quality factors on activities''~\cite{femmer2018quality}. We argue that the impact relationship can be more complex and requires a more general representation---i.e., rather than aiming for a taxonomy of impacts, we argue for developing an \textit{impact framework}.

Given the evaluation of quality factors on requirements entities on one side and the evaluation of activity attributes on the other side, the impact relationship between these variables can be formulated as a regression problem. Instead of relying on experts to hypothesize the (categorical) type or (linear) extent of an impact, more complex relationships can be determined using, for example, Bayesian data analysis~\cite{mcelreath2020statistical}. Consequently, this research stream aims to develop an impact framework capable of determining these impact relationships based on statistical instruments given sufficient data.

\subsection{Context Factors}
\label{sec:roadmap:context}
Context factors must be considered in the impact relationship to operationalize the requirements quality theory~\cite{mund2015does}. Large-scale endeavors acknowledge the importance of context factors in regard to requirements quality~\cite{fernandez2017naming}, yet no unified collection of context factors relevant to requirements engineering exists.
Established sets of software engineering context factors~\cite{petersen2009context,dybaa2012works} can be used as a starting point but require a dedicated investigation from the requirements engineering perspective.

A clear set of relevant context factors can support developing reporting guidelines for empirical studies on requirements quality and enable context-driven research~\cite{briand2017case}. While empirical software and requirements engineering publications typically strive for generalizability~\cite{dybaa2012works}, scoping an empirical study according to the given context factors allows the data collected in that study to be integrated into the impact framework as outlined in \Cref{sec:roadmap:impact}. Conversely, reporting the limited scope of a study enables a general requirements quality theory that can be assembled from multiple studies in well-defined contexts.

\subsection{Economic Impact}
\label{sec:roadmap:economics}

With the addition of economic concepts in the requirements quality theory, a research stream should be dedicated to the economic impact of activity facts. The impact relationship between quality factors and activities already benefits the acceptance of those factors for quality assurance in practice~\cite{femmer2018quality}. Adding an economic perspective---i.e., what amount of which resource a change of a certain activity-fact entails---can further bridge the gap between the normative, artifact-centric quality factors on one side and an economic decision-making process on the other side~\cite{deissenboeck2007economic}. Since the purpose of quality factors is to support quality assurance in industry, understanding this economic perspective is of high priority despite the complexity of the topic.

\subsection{Tool support}
\label{sec:roadmap:tool}

\begin{figure}
    \centering
    \includegraphics[width=\textwidth]{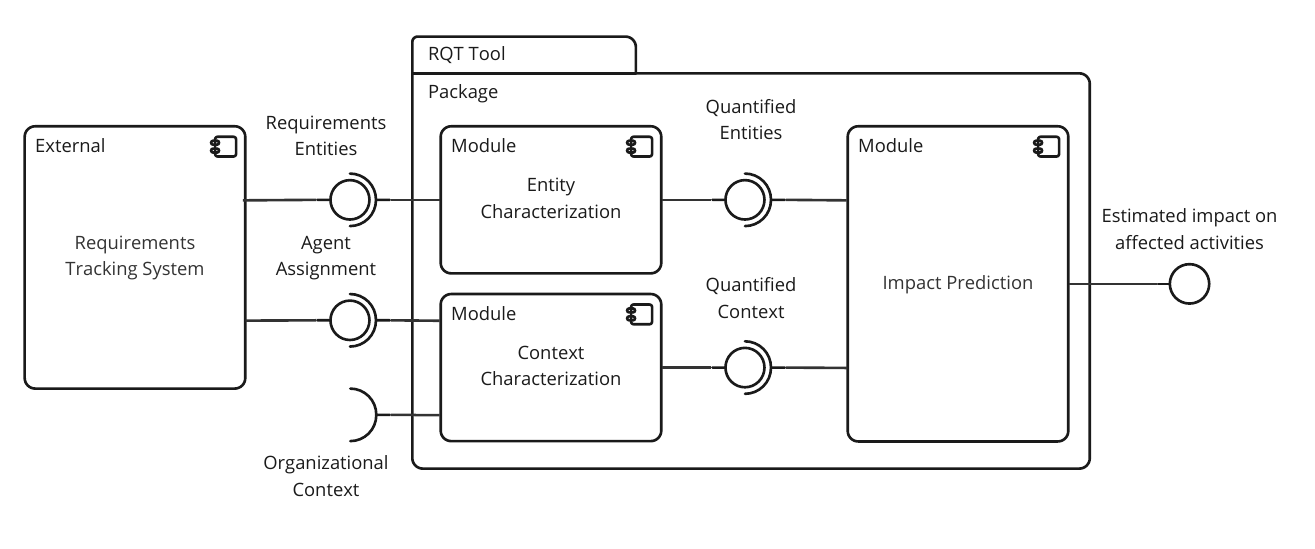}
    \caption{Architectural overview of the proposed tool-support}
    \label{fig:architecture}
\end{figure}

We aim to make the RQT applicable to the industrial context through the development of tool support. The components necessary to realize this tool support are visualized in \Cref{fig:architecture}. The goal of the tool is to estimate the impact of requirements entities and their context on the attributes of requirements-affected activities.

To this end, the tool needs an interface to the requirements entities, context information about the involved agents, and context information about the organization. The former two are often available in a requirements tracking system like Jira\footnote{\url{https://www.atlassian.com/software/jira}}~\cite{montgomery2022alternative}, while the latter a company likely has to generate and provide manually. 

Once provided with the necessary information, the tool characterizes both entities and context, i.e., quantifies the natural language requirements entities and the elusive factors determining the context. The quantified entities and context serve as input to the impact prediction model as described in \Cref{sec:roadmap:impact}, estimating the impact on the attributes of the requirements-affected activities, which in turn enables quantifying the economic impact as described in \Cref{sec:roadmap:economics}.

The realization of this tool depends on the previously described streams of research to identify valid quality factors (\Cref{sec:roadmap:factors}), context factors (\Cref{sec:roadmap:context}), and activity attributes (\Cref{sec:roadmap:artifacts}). For the tool to provide an automated impact prediction the following automation modules must be realized:

\begin{enumerate}
    \item Automatic entity characterization: a shared architecture to automatically evaluate the requirements quality factors collected in the quality factor ontology~\cite{frattini2022live}
    \item Automatic impact prediction: an accessible statistical model estimating the impact of quantified entities and context on affected activities, trained on historical data.
\end{enumerate}

Developing this tool while adhering to open science principles will allow scholars to propose new quality and context factors, customize relevant activity attributes, and contribute historic data to improve the impact estimation of the prediction model. We invite contributions to the implementation and maintenance of the tool via its dedicated repository on Github\footnote{Available at \url{https://github.com/JulianFrattini/rqt-tool}. An archived version is accessible at \url{https://doi.org/10.5281/zenodo.8167541}.}.

\section{Conclusion}
\label{sec:conclusion}

In this manuscript, we investigated the software quality literature and the application of the activity-based quality perspective to the requirements engineering domain. We extend the work of Femmer et al.~\cite{femmer2015ABREQM} by proposing an evolved and harmonized requirements quality theory, and assess the adherence of the requirements quality literature to this theory. Our survey confirms the bias towards artifact-centric and the negligence of activity-centric concepts, which was noted in previous secondary studies~\cite{montgomery2021empirical,frattini2022live}. Finally, we update the requirements quality research roadmap initiated by Femmer et al.~\cite{femmer2018quality} to guide future contributions in the requirements quality research domain.

We are confident that the harmonized requirements quality theory provides the necessary guidance to propel requirements quality research and establish a common understanding of quality that is operationalizable in practice. We invite fellow researchers to contribute to the theory and the requirements quality research field in adherence to it.

\bmhead{Acknowledgments}
This work was supported by the KKS foundation through the S.E.R.T. Research Profile project at Blekinge Institute of Technology.










\bigskip





\begin{appendices}



\end{appendices}

\bibliography{sn-bibliography}

\end{document}